\documentclass[twocolumn,noshowpacs,showkeys,amsmath,amssymb,preprintnumbers,nofootinbib,floatfix,pra]{revtex4}

\usepackage{graphicx}
\usepackage{dcolumn}
\usepackage{bm}
\usepackage{mathrsfs}
\usepackage{amsmath}
\usepackage{feynmf}
\usepackage{verbatim}
\usepackage{amssymb}

\begin{document}

\preprint{\footnotesize  SPIE Quantum Information and Computation VII, Vol. 7342-21, 73420M (2009)}
\title{Quantum algorithm for Bose-Einstein condensate quantum fluid dynamics: \\Twisting of filamentary vortex solitons demarcated by fast Poincar\'e recursion }

\author{Jeffrey Yepez${}^{1}$\thanks{To whom correspondence should be addressed.}, George Vahala${}^{2}$, Linda Vahala${}^{3}$}


\address{${}^{1}$Air Force Research Laboratory, Hanscom Air Force Base, Massachusetts  01731 \\
${}^{2}$Department of Physics, William \& Mary, Williamsburg, Virginia 23185\\
${}^{3}$Department of Electrical \& Computer Engineering, Old Dominion University, Norfolk, VA 23529}

\begin{abstract}
The dynamics of vortex solitons in a BEC superfluid  is studied.
A  quantum lattice-gas algorithm (localization-based quantum computation) is employed to examine the dynamical behavior of vortex soliton solutions of the Gross-Pitaevskii equation ($\phi^4$ interaction nonlinear Schroedinger equation).     Quantum turbulence is studied in large grid numerical simulations:  Kolmogorov spectrum associated with a Richardson energy cascade occurs on large flow scales.  At intermediate scales  a $k^{-6}$ power law emerges, in a classical-quantum transition from vortex filament reconnections to Kelvin wave-acoustic wave coupling.  The spontaneous exchange of intermediate vortex rings is observed. Finally, at very small spatial scales a $k^{-3}$ power law emerges, characterizing fluid dynamics occurring within the scale size of the vortex cores themselves, a characteristic Kelvin wave cascade region.   Poincar\'e recurrence is studied:  in the free non-interacting system, a fast Poincar\'e recurrence occurs for regular arrays of line vortices.  The recurrence period is used to demarcate dynamics driving the nonlinear quantum fluid towards turbulence, since fast recurrence is an approximate symmetry of the nonlinear quantum fluid at early times.  This class of quantum algorithms is useful for studying BEC superfluid dynamics over a broad range of wave numbers, from quantum flow to a pseudo-classical inviscid flow regime to a Kolmogorov inertial subrange.
\end{abstract}



\maketitle


\section{Introduction}

Quantum computation is a newly emerging technique within the field of information processing.  Feynman's conjecture  \cite{feynman-82}, now over a quarter century old, that a quantum mechanical computer could efficiently simulate physics, principally a many-body quantum system, has not yet been physically demonstrated on a quantum computer.  However, quantum algorithms (designed for future quantum computers) run today on supercomputers do provide a new numerical paradigm for performing stable and faithful simulations of quantum systems.  Here we use a quantum lattice-gas algorithm to model a Bose-Einstein condensate (BEC) superfluid in the notoriously difficult turbulent regime.  The quantum lattice-gas algorithm is strictly unitary, no dissipative terms are added to the reversible dynamics.  Thus, the quantum turbulence observed in the numerical simulation is rigorously obtained.
 
 Someday large-scale quantum simulators should provide experimental realization of the Feynman conjecture.
 Until such a time, one can learn much from qubit representations  of local dynamics occurring in a quantum simulator, particularly from the simplest qubit models (perhaps even bootstrapping toward a viable quantum computing architecture).    In this Letter, we employ a simple quantum computational model of a BEC superfluid, a minimal model of vortex-vortex ({\it e.g.} reconnection) and vortex-phonon interactions.  Its unitary representation is different than the Bose-Hubbard model \cite{PhysRevLett.81.3108}, which has served as the basis of quantum Monte Carlo simulations of trapped BEC superfluids \cite{PhysRevA.70.053615}.  No low-order Hamiltonian, such as the Bose-Hubbard Hamiltonian accounting for particle hopping and particle-particle interactions, is specified a priori to generate the dynamics.  Instead, exact dynamics is represented by basic unitary logic gates acting at the small lattice cell-size scale.  An effective Hamiltonian dynamics naturally emerges at large scales.  Thus the quantum lattice-gas model tested here is a high-energy model of the quantum particle dynamics that effectively and correctly behaves as a BEC superfluid on the large (low-energy) scale.  Large grids are employed to obtain good numerical predictions, typically $1024^3$ or larger, although accurate numerical results can be obtained in some cases for grid size an order of magnitude smaller.
 
  Previously, we demonstrated the practicality of our quantum lattice-gas computational model by exploring instabilities of soliton wave trains in 2+1 dimensions governed by a nonlinear Schr\"odinger wave equation \cite{yepez-vahala-qip-05}. 
Considering vortex solitons in BEC superfluids, we now focus on the behavior of vortex lattices as a means to comprehend quantum turbulence in 3+1 dimensions.   We use the small vortex lattices (quadrupolar arrangements of linear vortex solitons) as the basis of initial conditions to study the subsequent behavior a BEC superfluid flow.  These vortex lattices have zero total angular momentum, in contradistinction to quantized vortex lattices with fixed nonzero angular rotation in one direction that have been used as a marker of superfluid flow within a normal fluid background \cite{schunck:050404}.

The findings reported in this Letter are:
\begin{enumerate}
\item
Vortex lattices with multiple orthogonal angular rotation directions evolve into highly tangled turbulent like flows.  Energy spectra is consistent with a $k^{-5/3}$ Kolmogorov power law for vortex-vortex flow morphology for small wave numbers.

\item At large wave  numbers, the energy spectra is consistent with a $k^{-3}$ power law.

\item There is a well-defined cross-over region with its own characteristic power-law behavior.
  
\item
In free quantum fluids, or BEC superfluids with very weak nonlinear coupling, turbulent flow with tangled vortex filaments can quickly emerge but then just as quickly untangle.  Thus, fast Poicar\'e recurrence occurs in a weak BEC superfluid, albeit with remnant Kelvin wave (twisted vortex solitons).  

\item
A highly nonlinear mechanism for vortex-vortex interaction by exchanging vortex rings occurs in a BEC superfluid but also occurs in a free quantum fluid (linear Schroedinger wave equation).  

\end{enumerate}

\subsection{Application of measurement-based quantum computing}

There are different quantum informational representations of quantum computation including the standard quantum circuit model   and adiabatic quantum computing.  
Type-II quantum lattice gas models use state localization to boost the usefulness of quantum algorithms for physical simulation \cite{yepez-ijtp98,yepez-lncs99-short,yepez-ijmpc00a-short}.
Furthermore, type-II quantum computational models were the first examples of ``one-way" measurement-based quantum computing relying on a parallel array of locally entangled cluster states, an antecedent to cluster-based quantum computation 
\cite{gross:220503,nest:150504,childs:032318,PhysRevA.70.062314}.
 
  We employ a representation that relies on 2-qubit state localization ``self-steering" the outcome of the quantum computation.  Remarkably this mechanism gives rise to quantum fluidic behavior at the large scale,  {\it i.e.} the nonlinear interaction term in the Gross Pitaevskii equation, 
  the effective equation of motion of a low-temperature BEC superfluid.  With this representation, previously we have predicted solutions to a number of nonlinear classical and quantum systems \cite{PhysRevE.63.046702,yepez-berman-ezhov-kamenev-pra2002,vahala-yepez-pla03,vahala-yepez-pt04,yepez-vahala-qip-05}.  An advantage of our approach over the standard computational physics GP solvers is that 
  the simple unitary  collide-stream-rotate operations give rise to an algorithm that approaches pseudo-spectral accuracy \cite{yepez-cpc01} in much the same way as the simple collide-stream steps of a lattice Botlzmann algorithm approach pseudo-spectral accuracy for fluid dynamics simulations \cite{palpacelli:036712_short,succi-2001}.

\subsection{Overview of the modeling approach}


	Quantum fields are discretized on a cubic lattice.  Located at each lattice site is a local cluster state, in the simplest case just a single pair of qubits. The excited state, denoted by logical ``1", of a qubit $|q(\vec x_n,t)\rangle$ encodes a particle at the lattice site $\vec x_n$ at time $t$.  The local ket in the Fock basis of the qubit pair is
 \begin{equation}
\label{local_ket}
|\psi\rangle=\sum_{q,q'=0}^1\psi_{qq'} |q q'\rangle=\psi_0 |00\rangle + \psi_\uparrow |01\rangle  + \psi_\downarrow |10\rangle  + \psi_{\uparrow\downarrow} |11\rangle.
\end{equation}
The local Fock states $|01\rangle$ and $|10\rangle$ encode the spin states $\mid\uparrow\rangle$ and $\mid\downarrow\rangle$ and so
 the arrows as subscripts denote the respective amplitudes of those states.  The mean-field tensor product  $|\Psi(t)\rangle= \bigotimes_{n=1}^{L^3} |\psi(\vec x_n, t)\rangle$, where $L$ is the (linear) lattice size, represents the state of the quantum system.  Furthermore, to analytically recover the GP equation it is sufficient to consider just  the one-body sector, hence the 2-component field 
 \begin{equation}
\label{psi_2_component_form}
\psi(x) = \begin{pmatrix}
\psi_\uparrow(x)   \\
 \psi_\downarrow(x)  
\end{pmatrix}
\end{equation}
  is computed on the lattice.  The reduction from (\ref{local_ket}) to (\ref{psi_2_component_form}) is not an approximation.  (\ref{psi_2_component_form}) is an exact representation of the quantum dynamics so long as the simulation is carried out in the one-body sector.  Finally, the BEC wave function $\phi$  is determined as the sum of the amplitudes $\phi\equiv \psi_\uparrow+\psi_\downarrow$.  This, of course, retains the important effect of quantum interference.


	The evolution of $\psi$ is determined by $\uparrow$--$\downarrow$ qubit-qubit interactions (collide),  free motion of the amplitudes along the cubic lattice (stream), and qubit phase shifts to model the well known ``Mexican hat" interaction potential (rotation).  The qubit-qubit collisions are generated by a collision Hamiltonian $H_\text{c} = J\sigma_x$, where $J$ is the $\uparrow$--$\downarrow$ coupling energy, $\sigma_x$ a Pauli spin operator, and the collision time $\tau$ corresponds to the quantum logic gate time, chosen such that $\frac{J \tau}{\hbar}=\frac{\pi}{4}$. Free streaming of qubit states on the lattice  (emulating the motion of particles in space) is generated by the stream Hamiltonian $H_\text{s}=-i\hbar \sum_\text{lattice}\sigma_z \vec c\cdot \nabla$ acting on the qubits.   $\vec c$ is a streaming velocity along lattice directions. 
	
	Exploiting the fact that  $[H_\text{s}, H_\text{c}]\ne 0$, we use an interleaved compositional product of 2-qubit quantum gates generated $H_\text{s}$ and $H_\text{c}$ whereby an effective Hamiltonian  emerges in the scaling limit modeling the non-relativistic free particle Hamiltonian $H_\circ$.  Formally denoting this compositional product by the symbol $\circ$, the Hamiltonian  $H_\circ \equiv\text{Tr}[ H_\text{s} \circ H_\text{c}] \rightarrow -\frac{\hbar^2}{2m}\nabla^2+\cdots$, where in our model $m=\frac{1}{2}$ in lattice units and where the right arrow denotes a scaling-limit mapping.  Thus $H_\circ$ generates the evolution of a free scalar quantum field.  If this does not seem clear at the moment, do not fret because an explanation is presented in the section following immediately below.
	
	Phase rotation, inducing nonlinear particle-particle interactions,  is generated by $H_\text{int}(|\phi|^2)=\left(g|\phi|^{2}-1\right)|\phi|^{2}$, where the quantum logic gate $\tau'$ is chosen such that $\frac{g \tau'}{\hbar}\lessapprox 1$.   In the low energy limit, the local evolution is effectively
%
\begin{equation}
\label{quantum_equation}
\begin{split}
\phi(t+ 1, \vec x) = e^{ -i  H_\circ \tau/\hbar}\,e^{-i H_\text{int}\tau'/\hbar}\,\phi(t,\vec x)+\cdots,
\end{split}
\end{equation}
modeling (\ref{Gross_Pitaevskii_equation}) in the lowest order fluctuations.
A derivation of 
$H_\text{\tiny GP}\sim H_\circ + H_\text{int}$ presented in  \cite{yepez-epj-08}  follows from the typical scaling argument used in kinetic lattice gases, which we also review below.

\section{\label{lattice_gas_algorithm_Schroedinger_3D} Quantum lattice-gas algorithm}

Conservative two-qubit quantum gates have the form \cite{PhysRevE.63.046702}
\begin{equation}
\label{conservative_quantum_gate}
U_\text{\tiny conservative} |\psi\rangle =
\begin{pmatrix}
    1  & 0 &0&0   \\
    0 & {\cal A} & {\cal B} & 0 \\
    0 & {\cal C} & {\cal D} & 0\\
 0     &  0&0& {\cal E}
\end{pmatrix}
\begin{pmatrix}
      \psi_0   \\
      \psi_\uparrow\\
      \psi_\downarrow\\
      \psi_{\uparrow\downarrow}
\end{pmatrix},
\end{equation}
where the components in the zero-quantum subspace are a member of SU(2), {\it viz.} $\begin{pmatrix}
   {\cal A} & {\cal B}    \\
   {\cal C} & {\cal D}
\end{pmatrix}\in$ SU(2).
Only the local 2-component field's complex amplitudes $\psi_\uparrow(x)$ and $\psi_\downarrow(x)$ are quantum mechanically entangled by the action of (\ref{conservative_quantum_gate}).  Particle motion and particle-particle interactions are faithfully emulated strictly using quantum logic gates of the form of  (\ref{conservative_quantum_gate}).  Furthermore, to describe the quantum lattice-gas algorithm, it is sufficient to consider only the single-particle sector of the full quantum Hilbert space.   In this way, the algorithmic treatment becomes straightforward to describe using only $2\times 2$ matrices that represent the SU(2) subspace of (\ref{conservative_quantum_gate}).  

The justification for  this reduction is given in Ref.~\cite{PhysRevE.63.046702}. The quantum gate dynamics conserves particle number and consequently the effective $H_\text{\tiny GP}$ in (\ref{quantum_equation}) commutes with the particle number operator.  Thus, $H_\text{\tiny GP}$  is block diagonal over the $n$-body sectors of the Hilbert space.  One is justified to run a simulation in any one of the $n$-body sector, for $0\le n \le 2 L^3$, which is an exact representation of all the relevant quantum dynamics in that sector, independent of the dynamics in all the other sectors.  The advantage of limiting the simulation to the one-body sector  is that the algorithmic complexity scales linearly with the number of qubits and the one-body sector is sufficient to capture all the relevant physics in the BEC.  It is for this reason that the type-II quantum algorithm can be implemented on a classical supercomputer while retaining its exactness as a quantum mechanical simulation.

We now describe the quantum algorithm by dealing with the zero-quantum subspace part $\begin{pmatrix}
{\cal A} & {\cal B} \\
 {\cal C} & {\cal D}
\end{pmatrix}$ of the quantum logic gate operators (\ref{conservative_quantum_gate}).
 The SU(2) quantum operator $C= e^{i   \frac{\pi}{4} \sigma_x (1-\sigma_x)}$
%
%
that acts locally at every point $x$ by the map 
\begin{equation}
\label{collision_map}
\text{local collision}: \qquad \psi(x) \rightarrow C \,\psi(x).
\end{equation}
The well known Pauli matrices are
%
\(
{\scriptsize
\sigma_x =\begin{pmatrix}
   0   & 1   \\
   1   &  0
\end{pmatrix}
}
\), 
\(
{\scriptsize
\sigma_y =\begin{pmatrix}
   0   & -i   \\
   i   &  0
\end{pmatrix}
}
\), 
\(
{\scriptsize
\sigma_z =\begin{pmatrix}
   1   & 0   \\
   0   &  -1
\end{pmatrix}.
}
\)
%
The complex scalar density $\phi = (1,1) \cdot \psi
= \psi_\uparrow + \psi_\downarrow$ is conserved by (\ref{collision_map}), and consequently the probability $|\psi|^2$ is also conserved locally. Local equilibrium ({\it i.e.} $\psi = C \,\psi$) occurs when the amplitudes are equal ($\psi_\uparrow=\psi_\downarrow$), but in general such a local equilibrium is then broken if a spinor component 
is displaced in space by the vectorial amount $\Delta \vec x$.  To conserve probability, we admit only complementary displacements of the field components, induced by the stream operators of the form
\begin{subequations}
\begin{equation}
\label{stream_operators}
S_{\Delta\vec x, 0} = n + e^{\Delta \vec x \partial_{\vec x}}\,\bar n,
\qquad
 S_{\Delta\vec x, 1} =\bar n + e^{\Delta \vec x \partial_{\vec x}} \,n,
 \end{equation}
\end{subequations}
where $n=\frac{1}{2}(1-\sigma_z)$ and $\bar n=\frac{1}{2}(1+\sigma_z)$. 
Although the application of (\ref{stream_operators}) usually breaks local equilibrium induced by (\ref{collision_map}), with the appropriate boundary conditions, for example  periodic boundary conditions, (\ref{stream_operators}) is guaranteed to conserve the total density $\int d^3x\, \psi(\vec x)$, and in turn the total probability $\int d^3x \,\left(|\psi_\uparrow(\vec x)|^2+|\psi_\downarrow(\vec x)|^2\right)$.  To construct a quantum algorithm using a combination of the operators $C$ and $S_{\Delta \vec x,\gamma}$, for $\gamma=0$ or $1$, and the respective adjoints, we restrict our considerations to those combinations which are close to the identity operator.   Our basic approach uses  
the interleaved operator
\begin{equation}
\label{interleaved_operator}
I_{x\gamma} =  S_{-\Delta \vec x,\gamma}  C^\dagger  S_{\Delta \vec x,\gamma}  C
\end{equation}
as the basic building block of the quantum algorithm.
For example, an evolution operator for the $\gamma$th component is 
 \begin{equation}
\label{basic_typeII_quantum_algorithm}
  U_ \gamma[\Omega(\vec x)]= I_{x\gamma}^2 I_{y \gamma}^2 I_{z \gamma}^2 e^{-i  \varepsilon^2\Omega(\vec x)} ,
\end{equation}
where $\varepsilon \sim \frac{1}{N}$, where $N$ is the grid resolution ({\it i.e.} $N$  is the number of grid points along one edge of the simulation volume).  (\ref{basic_typeII_quantum_algorithm}) represents the three aspects of a type-II quantum algorithm: stream, collide, and state reduction. In dimensionless units ($c=1$), note that $\varepsilon^2\sim \Delta x^2 \sim \Delta t$.
 This evolution operator is spatially dependent only through local state reduction $\Omega$:
 \begin{equation}
\label{basic_unitary_evolution_equation}
\psi(\vec x, t+\Delta t) =  U_ \gamma[\Omega] \,\psi(\vec x, t).
\end{equation}
(\ref{basic_unitary_evolution_equation}) specifies the nonlinear quantum lattice gas model.

\subsection{Effective field theory}

The R.H.S. of (\ref{basic_unitary_evolution_equation}) can be expressed as a finite-difference of $\psi$.  Taylor expanding the R.H.S. in $\varepsilon$, one obtains the following quantum lattice gas equation
\begin{equation}
\label{quantum_lattice_gas_equation_spinor_form}
\begin{split}
\psi(\vec x, t+\Delta t) = 
\psi(\vec x, t)
-i \varepsilon^2 \left[
-\frac{1}{2}\gamma_x\nabla^2
+\Omega 
\right]
\psi(\vec x, t)
\\
+
\frac{(-1)^\gamma\varepsilon^3}{4}(\sigma_y+\sigma_z)\nabla^3
\psi(\vec x, t)
+
{\cal O}(\varepsilon^4),
\end{split}
\end{equation}
 where $\gamma=0$ or $1$.
Since the order $\varepsilon^3$ error term in (\ref{quantum_lattice_gas_equation_spinor_form}) changes sign with $\gamma$, we can induce a cancelation of this error term using a symmetric evolution operator 
%
%
\begin{equation}
\label{symmetrized_evolution}
U[\Omega] = U_{1}\left[\frac{\Omega}{2}\right]U_{0}\left[\frac{\Omega}{2}\right]
\end{equation}
%
%
that is invariant under field component interchange.
Therefore, the following quantum map $\psi(\vec x,t+\Delta t) =  U[\Omega(\vec x)] \psi(\vec x, t)$
 leads to the quantum lattice gas equation
 
 %
\begin{equation}
\label{quantum_lattice_gas_equation_improved}
\psi(\vec x, t+\Delta t) = 
\psi(\vec x, t)
-i \varepsilon^2 
\left[
-\frac{1}{2}\sigma_x\nabla^2
+\Omega 
\right]
\psi(\vec x, t)
+
{\cal O}(\varepsilon^4).
\end{equation}
Now, in the low energy scaling limit,  we have $\frac{1}{\varepsilon^2}\left[
\psi(\vec x, t+\Delta t) - 
\psi(\vec x, t)
\right]\rightarrow \partial_t\psi(\vec x, t)$.  Therefore,   dividing both sides of (\ref{quantum_lattice_gas_equation_improved}) by $\varepsilon^2$, 
 the quantum lattice gas equation is
\begin{equation}
i \partial_t \psi = 
- \sigma_x\nabla^2 \psi 
+
 \Omega \psi +
{\cal O}(\varepsilon^2)
\end{equation}
 in the low energy and low momentum limit.
Finally, since $\phi =\alpha+\beta$,   
taking the density moment gives the effective scalar field equation
\begin{equation}
\label{Schroedinger_effective_field_theory_improved}
i \partial_t \phi = 
-\nabla^2 \phi 
+
 \Omega\, \phi +
{\cal O}(\varepsilon^2),
\end{equation}
which is the Schroedinger wave equation with  $m=\frac{1}{2}$ for $\hbar=1$, so long as $|\Delta \vec x|^2=\Delta t=\varepsilon$.    From the order of the error term in (\ref{Schroedinger_effective_field_theory_improved}), the Taylor expansion predicts that the quantum algorithm is second order convergent in space.

The low energy effective Hamiltonian that is the model generator of the evolution, $U = e^{i \Delta t H_\text{eff}/\hbar}$, is the following
\begin{equation}
\label{H_effective}
H_\text{eff} = -\frac{\hbar^2}{2m} \nabla^2 +\hbar \Omega(\vec x)+ {\cal O}(\Delta t, \Delta x^2),
\end{equation}
  where we have written the quantum diffusion coefficient as $\frac{\Delta x^2}{\Delta t} = \frac{\hbar}{m}.$
This is the nonlinear GP Hamiltonian since
$\hbar \,\Omega(\vec x) =  g |\phi(\vec x)|^2-1$, where $g$ the on-site interaction energy. 

\subsection{Matching BEC experiments}

BEC optical lattice setups tuned to a Feshbach resonance can simulate either repulsive or attractive $g=\pm\frac{4\pi\hbar^2}{a_\circ}$, with $a_\circ$  is the scattering length \cite{RevModPhys.71.463}.  In the simulations presented below a repulsive nonlinear interaction is used.  We have been careful not to choose $g$ too large, as (\ref{Gross_Pitaevskii_equation}) is appropriate for weak scattering;  a large $g$ triggers the occurrence of three-body molecular interactions, violating (\ref{Gross_Pitaevskii_equation}) and destroying the BEC.  In setups with  $\sim 10^5$ atoms, $g\le 100$ is accessible.  So a quantum lattice gas simulation is realistic so long as $g\le 100$.

\section{Vortex solitons}

\subsection{Single line soliton}

We first seek a line vortex steady state solution of the rescaled GP equation 
\begin{equation}
\label{rescaled_GP_equation}
- \frac{1}{a}\nabla^2 \phi + ( g | \phi |^2-1 )\phi=0.  
\end{equation}
We use a spatial rescaling parameter  $\sqrt{a}$ so that the vortex core can be resolved on the computational lattice with sufficient numerical accuracy. 
(\ref{rescaled_GP_equation}) is modeled by using $\Omega = g |\phi|^2-1$ in  (\ref{symmetrized_evolution}), and this corresponds to an interaction Lagrangian density ${\cal L}_\text{int}=a\,\phi^\ast\phi\left(1-g \phi^\ast\phi\right)$.  Then (\ref{H_effective})  becomes the nonlinear Hamiltonian for the GP equation (\ref{rescaled_GP_equation}).  With the appropriate choice of nonlinear coupling $g$ and normalization of the wave function, physically the  energy of a constant external potential can cancel the internal interaction energy, at least in the region of bulk flow.  This is the type of solution that I will explain here.

A solution for the wave function of the vortex soliton is found by separation of variables in polar coordinates.  Inserting $\phi(r,\theta)=f(r) e^{i n\theta}$, for integer $n$, into  (\ref{rescaled_GP_equation}) then gives
\begin{equation}
\frac{d^2 f(r)}{adr^2} +\frac{1}{a r}\frac{d f(r)}{dr}-\frac{n^2}{ar^2}f(r) + \left(1- g f(r)^2\right) f(r) =  0,
\end{equation}
which can be solved for any winding number.  For the simplest $n=1$ and $g = 1$ case, a Pad\'e approximant of the spatially scalable form is
\begin{equation}
\label{Pade_approximant}
f(r) = \sqrt{\frac{11 a r^2 (12 + a r^2)}{384+ a r^2 (128 + 11 a r^2)}}.
\end{equation}
Then, rescaling $\sqrt{a}\,r\rightarrow r$,  we have
\begin{equation}
\label{radial_equation}
R''( r) +\frac{1}{r}R'(r)-\frac{n^2}{r^2}R(r) + \left(1-R(r)^2\right) R(r) =  0,
\end{equation}
where $R(r)=f\left(\frac{r}{\sqrt{a}}\right)$ is the solution of the radial part of (\ref{Gross_Pitaevskii_equation}) found by Berloff \cite{berloff-jpamg-2004}.  Rescaling  allows one to arbitrarily resolve the vortex core on the computational lattice to achieve sufficient numerical accuracy.

(\ref{rescaled_GP_equation}) satisfies the free field condition in the bulk when $|\phi|^2 \rightarrow \frac{1}{g}$  (far away from any vortex singularity).  This is called {\it bulk point normalization}. That is, since
%
\(
\lim_{r\rightarrow \infty} f(r) =1,
\)
%
to satisfy the free field condition we must choose $g = 1$.
Thus with $\sqrt{a}\,r\rightarrow r$, at the end of the day the equation we are going to numerically solve is
\begin{equation}
i \partial_t \phi = - \nabla^2 \phi + ( | \phi |^2 -1 )\phi=0
\end{equation}
if we were to model a single vortex soliton. Unfortunately, in a box with periodic boundary conditions we cannot make do with just a single vortex soliton.  We need at least four of them.

\subsection{Arbitrary coupling strength via wave function normalization}

The GP equation is invariant under arbitrary wave function normalization provided one also rescales the nonlinear interaction:
 %
\begin{equation}
\label{Gross_Pitaevskii_equation_rescaled}
i \partial_t \phi = - \nabla^2 \phi + a( g | \phi |^2 -1)\phi\,
\end{equation} 
%
is invariant under $\phi \rightarrow  ({\alpha \,a^{\frac{N}{2}}L^\frac{3}{2}})^{-1}  \phi$ 
and $g \rightarrow  \alpha^2 a^N L^3 g$, 
where $a$ is the spatial scaling parameter in (\ref{Pade_approximant}) and $N$ is the number of vortex lines, {\it e.g.} $N=4$ in (\ref{4_vortex_solitons}), $L$ is the grid size, and the factor $\alpha = a^{-\frac{N}{2}}L^{-\frac{3}{2}}\left(\int d^3 x |\phi(\vec x) |^2\right)^{\frac{1}{2}} \lessapprox 1$ accounts for the excluded volume due to the vortex cores.

\subsection{Quadrupolar configuration (4 line solitons)}

\begin{figure}[tbhp]
\begin{center}
\includegraphics[width=3.4in]{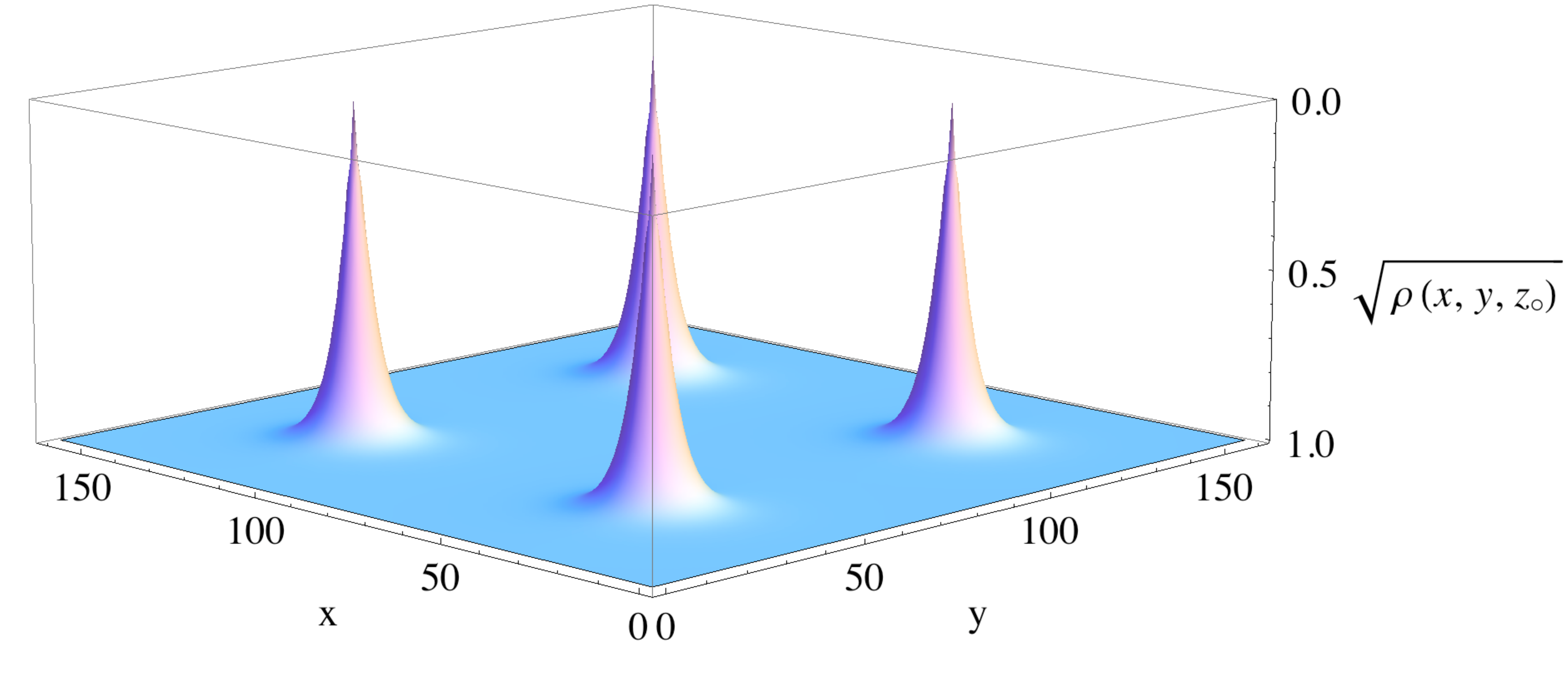}
\includegraphics[width=3.4in]{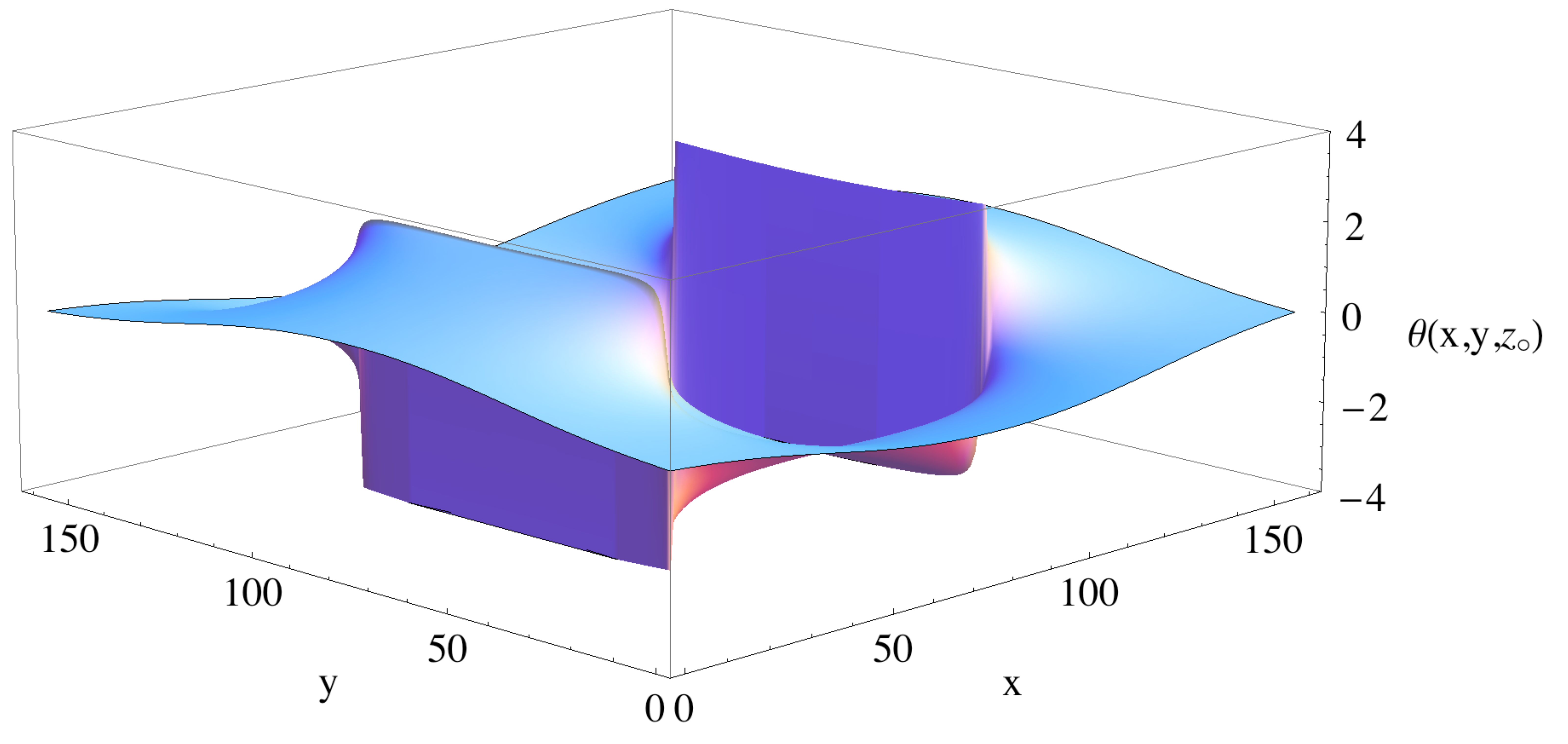}
\includegraphics[width=2.0in]{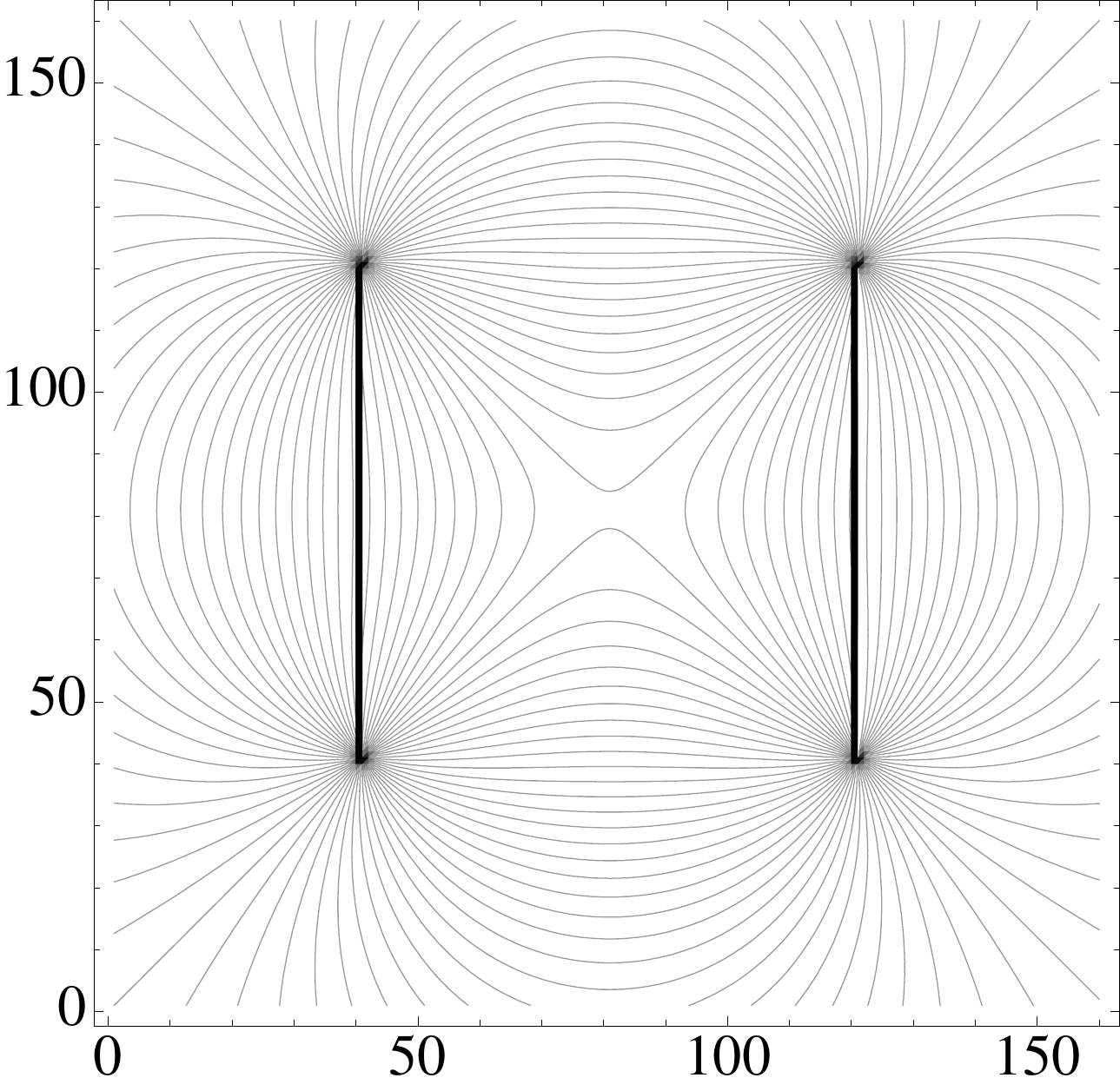}
\caption{\label{initial_conditions} \footnotesize A slice at $z=z_\circ$ of the magnitude $\sqrt{\rho(x,y,z_\circ)}$ (top, upside down) and phase $\theta(x,y,z_\circ)$ (middle) and phase contours (bottom) of the wave function for a vortex quadrupole, the product of 4 vortex soliton solutions on a grid of size $L=160$.  The density $\rho(x,y)=|\phi(x,y)|^2$ and so $\sqrt{\rho(x,y,z_\circ)}\rightarrow 1$ away from a vortex core (in the bulk). From the phase diagram, plotted $-\pi\le \theta(x,y,z_\circ)\le \pi$, going around any contour in the $z=z_\circ$ plane that encloses a single vortex singularity accumulates a phase of $\pm 2\pi$ radians.  With $N=4$ line vortices one can accommodate periodic boundary conditions in the phase.}
\end{center}
\end{figure}

A simple initial condition that ensures periodicity is  four symmetrically displaced vortex line solitons (parallel to the z-axis for the time being) in product form
\begin{subequations}
\label{4_vortex_solitons}
\begin{eqnarray}
\nonumber
\phi(x,y) 
&= &f(r_{++})e^{i \theta_{++}}\times f(r_{+-})e^{-i \theta_{+-}}\\
&\times& f(r_{-+})e^{-i \theta_{-+}}\times f(r_{--})e^{i \theta_{--}}\\
\nonumber
&= &f(r_{++})f(r_{+-})
 f(r_{-+})f(r_{--})
 \\
& \times& e^{i (\theta_{++}-\theta_{+-}-\theta_{-+}+\theta_{--})},
\end{eqnarray}
\end{subequations}
where the radial distance from a vortex line along the $z$-axis is 
\begin{subequations}
\begin{eqnarray}
r_{++}(x,y)  & = & (x-x_\circ + \delta)^2 + (y-y_\circ +\delta)^2 \\
r_{+-}(x,y)  & = & (x -x_\circ + \delta)^2 + (y-y_\circ-\delta)^2  \\
r_{-+}(x,y)  & = & (x -x_\circ - \delta)^2 + (y-y_\circ+\delta)^2 \\
r_{--}(x,y)  & = &(x -x_\circ - \delta)^2 + (y-y_\circ-\delta)^2.
\end{eqnarray}
\end{subequations}
The size of the vortex quadrupole is $|2\delta|$,  $\text{sign}(\delta)=\pm1$ is called the {\it polarity} of the vortex quadrupole, and its center is $(x_\circ,y_\circ)$.  The phase angles are 
\begin{subequations}
\begin{eqnarray}
\theta_{++}(x,y) & = &\arctan\frac{ y-y_\circ+\delta}{x-x_\circ- \delta}  \\
\theta_{+-}(x,y) & = & \arctan\frac{ y-y_\circ+\delta}{x-x_\circ- \delta} \\
\theta_{-+}(x,y)  & = & \arctan\frac{ y-y_\circ-\delta}{x-x_\circ+\delta}  \\
\theta_{--}(x,y)  & = & \arctan\frac{ y-y_\circ-\delta}{x-x_\circ-\delta}.
\end{eqnarray}
\end{subequations}
The magnitude and phase of (\ref{4_vortex_solitons}) are plotted in Fig.~\ref{initial_conditions} with $\delta = \frac{L}{4}$ and $a=0.1$ and $L=160$, demonstrating the periodicity of (\ref{4_vortex_solitons}). 
We shall use such quadrupole vortex lattice configurations aligned along orthogonal principal lattice direction to represent initial conditions for numerical simulations.

\section{Quantum turbulence}

Turbulence remains one of the greatest unsolved problems of classical physics, even in the incompressible limit.  
  The complexities of classical turbulence are further compounded by the inability to accurately define a classical ``eddy."  This creates ambiguity in the phenomenology of the inertial cascade \cite{PRSL_Richardson_1926} and the self-similarity in the dynamics of the smaller eddies that leads to the famous Kolmogorov $k^{-5/3}$ energy spectrum \cite{Kolmogorov-1941,PhysRevLett.31.744}.

Feynman proposed that the superfluid turbulent state consists of a tangle of quantized vortices \cite{Feynman.1955}.  
  	Quantum turbulent studies, both experimental \cite{parker:145301} and computational 
\cite{PhysRevLett.78.3896,kasamatsu:063616}, 
	are being pursued vigorously--not only for their intrinsic importance but also  in the hope of shedding light on classical turbulence.  
	 This is consistent with behavior observed in our quantum algorithms simulations of a BEC.
\begin{figure}[htbp]
\begin{center}
\includegraphics[width=3.4in]{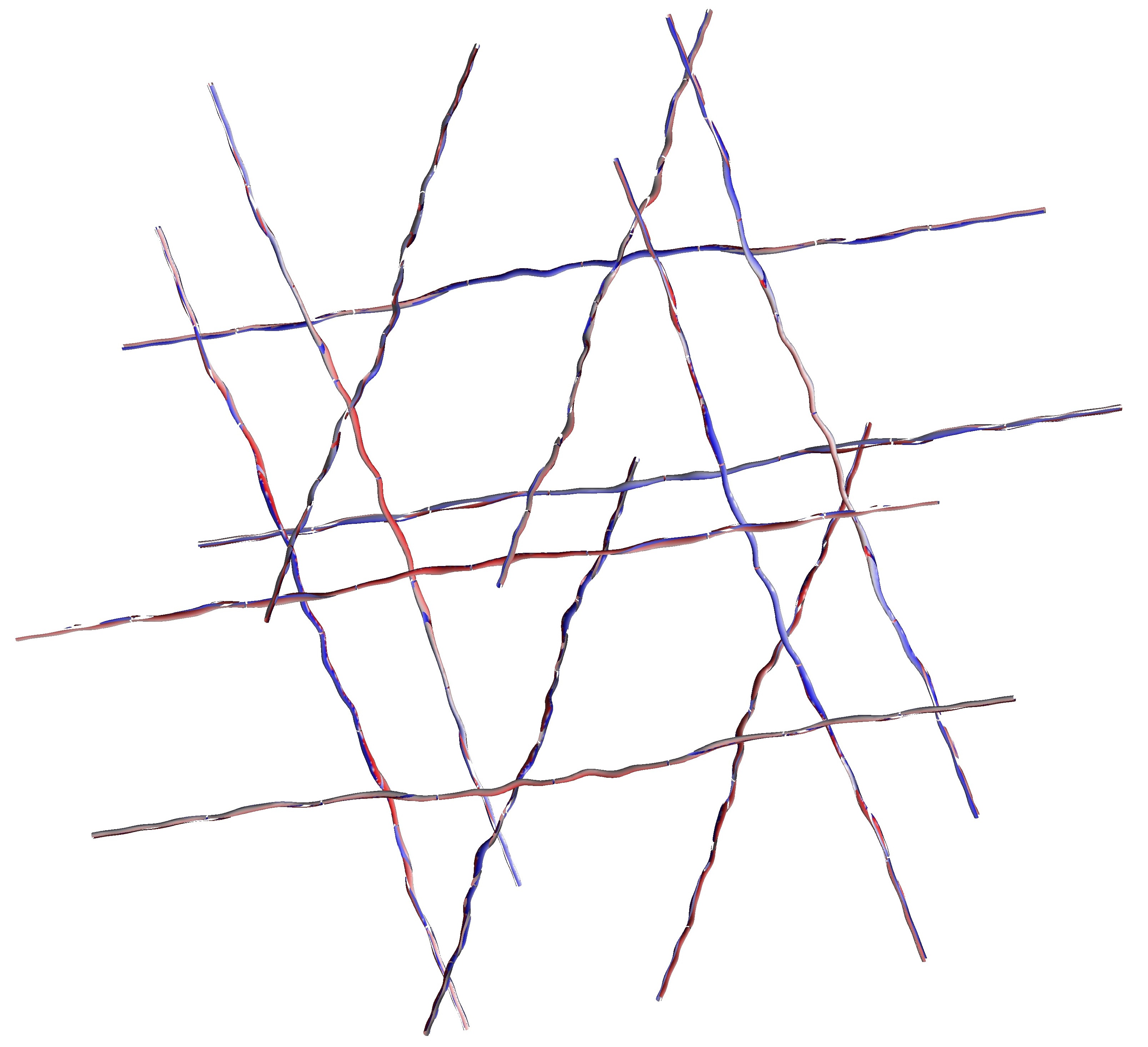}
\includegraphics[width=3.4in]{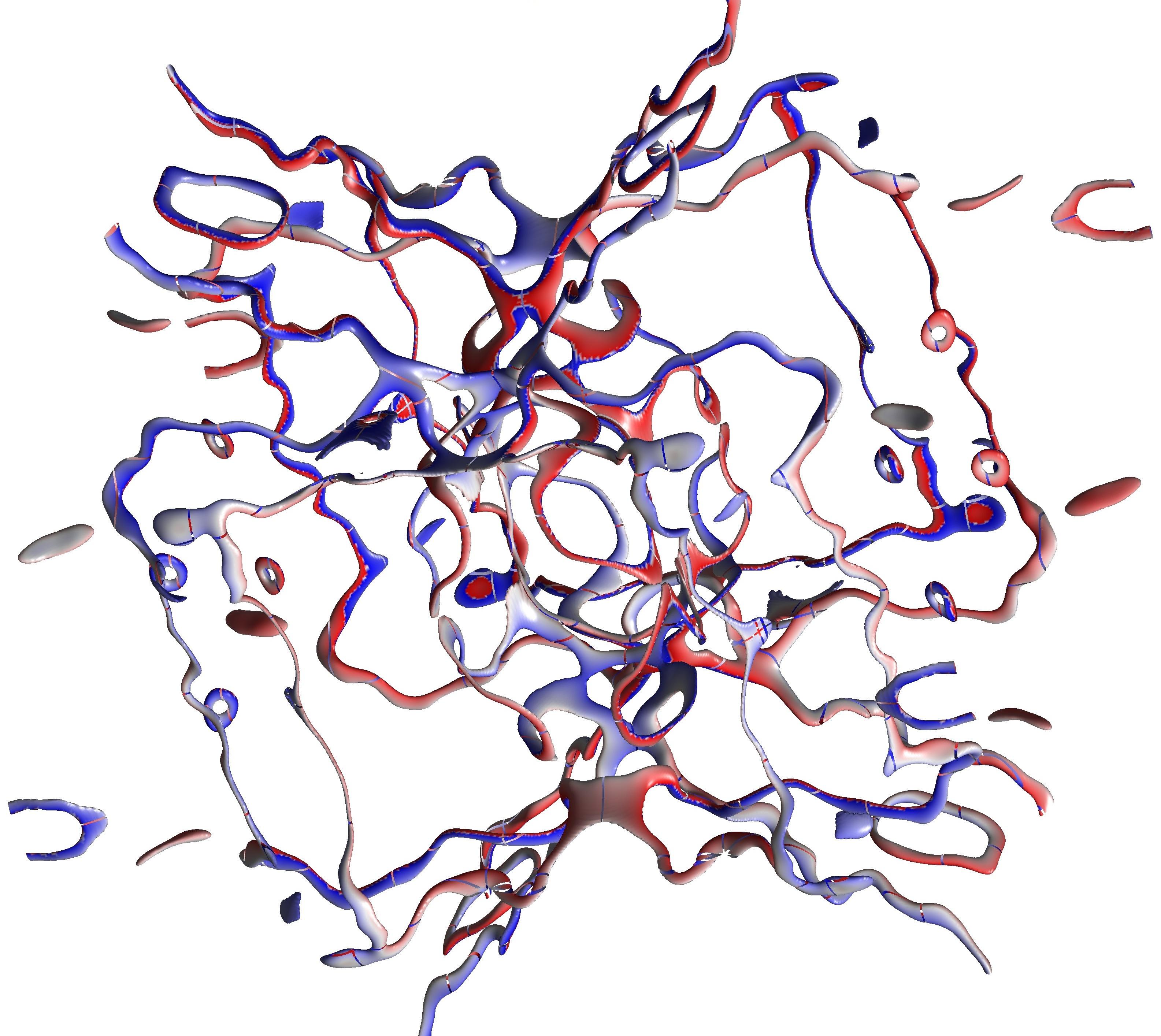}
\end{center}
\caption{\footnotesize  \label{hawk}
Starting with $N=12$ vortex lines on a $512^3$ lattice. Vortex tubes at $t=200K$ (top) show an onset of a Kelvin wave instability.  Tangled vortices are observed, even when $H_\text{int} \sim 0$, at  $t=3.3K$ (bottom).  
Remarkably, one observes many vortex rings mediating the vortex line-line interactions.
}
\end{figure}
In contrast to classical eddies, a quantized vortex is a well defined stable topological line defect in the phase of the complex scalar quantum wave function and a soliton in the magnitude of the wave function.  The vorticity is generated in a multiply-connected region where the circulation is quantized:  the phase of the complex scalar wave function along a contour  is accumulated in integer multiples of $2\pi$ for any contour enclosing a quantum vortex center.  Early superfluid turbulence studies focused on a regime describable by Landau's two-fluid equations that modeled the interaction between the inviscid superfluid and a viscous normal fluid \cite{PhysRev.60.356} and understood in terms of the motion of quantized vortices due to the Magnus force \cite{PhysRevB.44.9667}.  
In constrast, a BEC superfluid comprises a macroscopic number of integer spin particles in effectively a zero temperture ground state, and its dynamics below the critical temperature is remarkably well described by the   Gross-Pitaevskii (GP) equation 
\cite{JMathPhys.1963.4.195,JETP.1961.2.451}, 
which in normalized form is
\begin{equation}
\label{Gross_Pitaevskii_equation}
i \partial_t \phi = - \nabla^2 \phi + ( | \phi |^2 -1)\phi\,,
\end{equation} 
where $\phi$ is the mean-field BEC wave function.  It is well known \cite{nore:2644}
that the Madelung transformation $\phi =\sqrt{\rho} \,e^{ i \theta}$ on the GP equation results in compressible inviscid fluid equations for the density $\rho=|\phi|^2$ and velocity  $\mathbf{v} =2 \nabla \theta$, with the appearance of quantum pressure terms in the momentum and energy equations.

\subsection{Viewing quantum turbulence}

As mentioned above, products of quadrupolar line solitons depicted in Fig.~\ref{initial_conditions} may be shifted and rotated about any direction to build up unstable yet periodic initial conditions.   
Here we report on a simulation initialized with 3 orthogonal sets of 4 vortex centers (total of $N=12$ vortex lines) on a $512^3$ and $1024^3$ grids. 
For $N=12$ superfluid simulation shown in Fig.~\ref{hawk},  transverse vortex waves (helical perturbations of the vortex tube away from its cylindrical shape) known as Kelvin waves onset on the  filamentary core by $t=200$ time steps and subsequent become unstable, the initial drive force of the energy cascade \cite{PhysRevLett.86.3080} (large waves coupling to  smaller waves). 
  There is a rapid tangling of the quantized vortices (by $t\sim 3K$).  

One might expect a stark difference between the observed quantum turbulence arising from the compressible inviscid fluid equations of the GP equation without viscous shear dissipation, since unitary quantum dynamics is Hamiltonian, and classical turbulence arising from the incompressible Navier-Stokes equation with viscous shear dissipation, since viscous hydrodynamics system is non-Hamiltonian.   There are clear similarities for small wave numbers.

  \subsection{Spectrum for incompressible kinetic energy}

\begin{figure}[thbp]
\begin{center}
\includegraphics[width=3.4in]{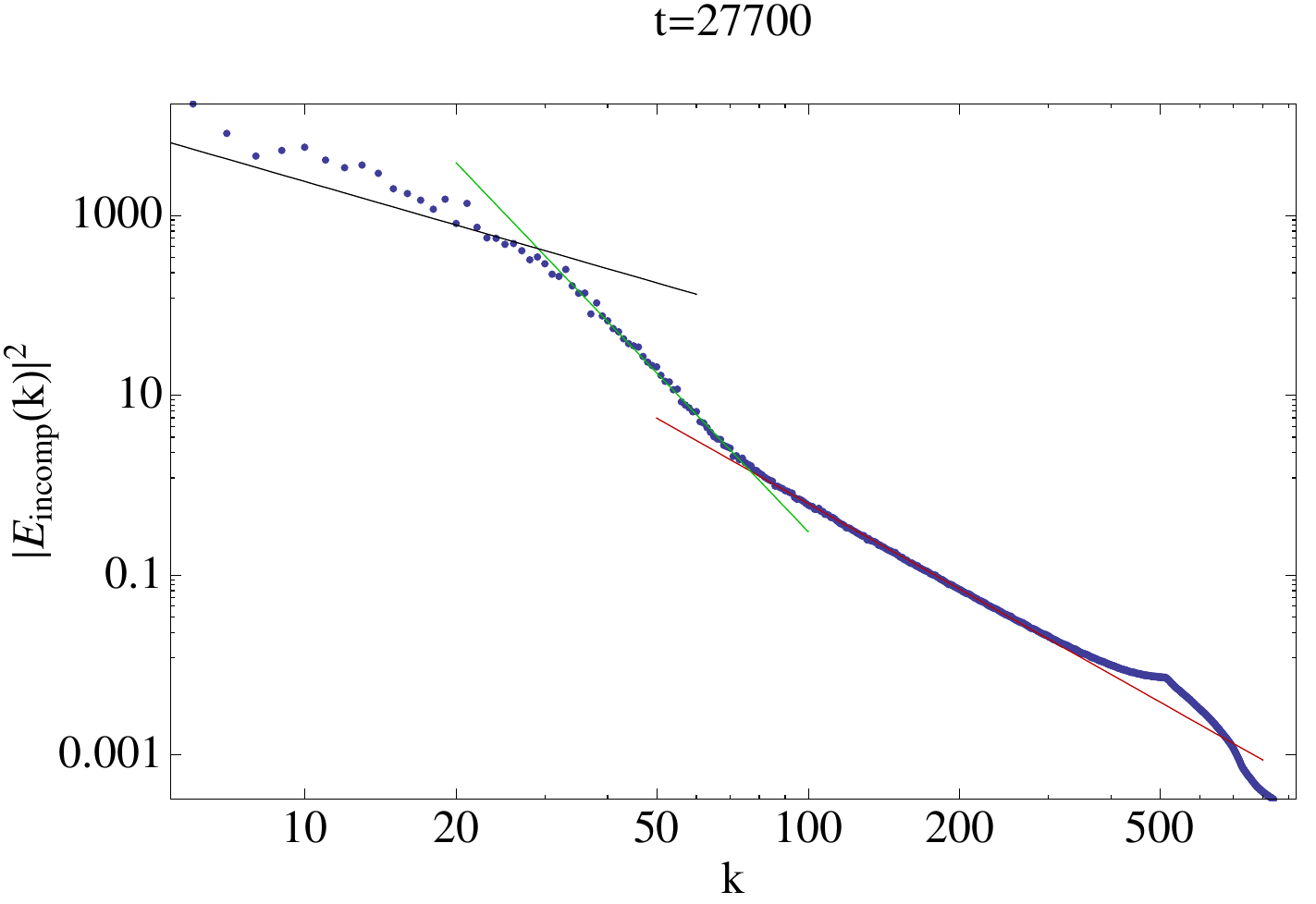}
\includegraphics[width=3.4in]{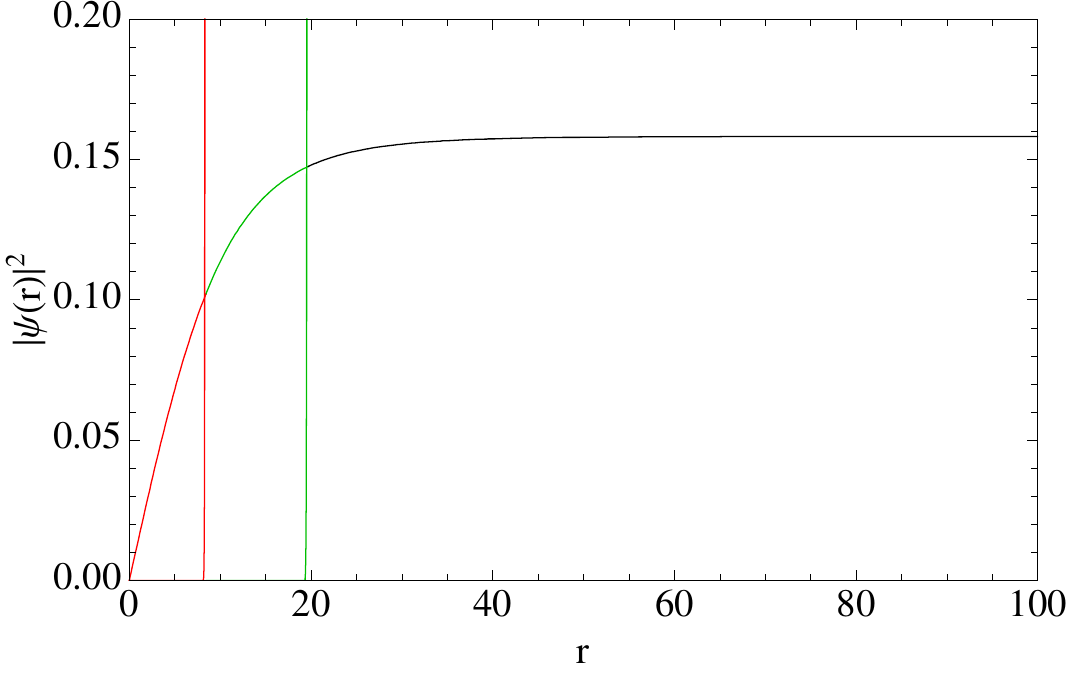}
\end{center}
\caption{\footnotesize  \label{spectrum}
Power spectrum  of the quantum fluid's incompressible kinetic energy (top). There are three regions characterized by differing power laws displayed on the vortex soliton spatial profile (bottom).  Numerical data (dots) is from a supercomputer simulation of a quantum lattice gas on a  $1024^3$ grid.  Kolmogorov (black), transition (green), and core interior (red) regimes are shown.}
\end{figure}

The approach we follow is to decompose the conserved energy into its kinetic, quantum and internal energy parts   \cite{nore:2644}, as follows $E_\text{tot}
= 
E_\text{kin}
+
E_\text{qu}
+
E_\text{int} = const.$
\begin{equation}
\label{energies}
\left(E_\text{kin}, E_\text{qu},E_\text{int}\middle)
=
\frac{1}{4}
\int d^3 x
\middle(
\left(
\sqrt{\rho}
\mathbf{v}
\right)^2
,
\left(
2 \nabla
\sqrt{\rho}
\right)^2
,
2  \rho^2
\right),
\end{equation}
  Vortex reconnections through Kelvin wave instabilities contribute to the Richardson cascade  \cite{PRSL_Richardson_1926}   (large  vortices break into smaller ones)  leading to the famous  Kolmogorov spectrum.  The presently held notion is that vortex filaments reduce down into smaller and smaller loops and the high-$k$ oscillations of the vortex center couple to elementary phonon excitations at the healing length giving rise to a quantum dissipation scale.  Vortices are destroyed by the emission of sound waves  (compressible excitations) that escape across the space.

  A Kolmogorov spectrum is observed in incompressible and quantum energies of the BEC superfluid for $k< 30$, see Fig.~\ref{spectrum}.  The emergence of full developed quantum turbulence is plotted at time $t=27.7K$ for a $1024^3$ lattice starting with $N=12$ vortex solitons (1 quadrupole per spatial direction).  The measured power law $k^{-1.61}$  (black) may suggest that the theoretical $k^{-\frac{5}{3}}$ Komolgorov power law describes the spectrum for $k\lesssim 20$, although the fit is not excellent.  A new power law $k^{-5.87}$  (green) emerges in a transition region from $30 \lesssim k\lesssim 70$.  After $k\gtrsim 70$, the observed power law $k^{-3.16}$ (red) agrees with the theoretical prediction of $k^{-3}$ and excellently fits the data in this region.  The power law fits were all computed using linear regression.

The bottom plot in Fig.~\ref{spectrum} shows the three power regimes in a spatial view by overlaying the cut-off lengths on the vortex core profile.  The $k^{-3}$ power law characterizes fluid dynamics within the vortex core itself, with an upper cut-off scale measured to occur at $r=8.44$ (in units of lattice sites).  The  $k^{-5.87}$ power law occurs near the boundary of the vortex core, with lower and upper spatial cut-offs, $8.44\lesssim r \lesssim 19.71$.  The $k^{-1.61}$ power law occurs on spatial scales larger than vortex core size.  For very large wave numbers $k\gtrsim 300$ (with a cut-off of about just 2 lattice cell sizes), the spectrum all together drops off the chart, as expected.

\section{Poincar\'e recurrence}

\begin{figure}[thbp!]
\begin{center}
\includegraphics[width=2.0in]{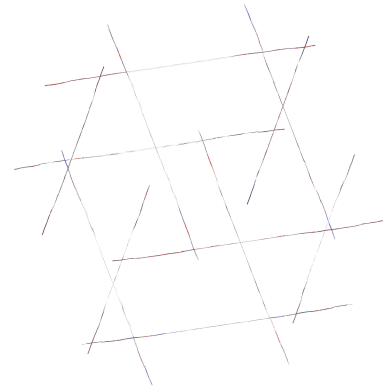}
\end{center}
\caption{\footnotesize  \label{hawk2}
In $N=12$ simulation, untangled vortices are observed $t=21K$ (bottom).  
The inital state recurs after a turbulent state. 
An ordered state at $t=21K$ deterministically returns to the initial state untangling turbulence, a cascade that cycles at intervals of $t_\text{\tiny P}=21K$ on a $512^3$ lattice.
}
\end{figure}
In Hamiltonian systems, the dynamics must be invertible, so it is possible to observe a reverse cascade, and on time scales shorter than otherwise expected;  vortex tubes untangle and reform by the absorption of sound waves to recover a configuration close to a configuration that occurred earlier in time in the flow.  
Recursion arising from the Hamiltonian system is observed in animations of the flow.
This occurs in the limit of vanishing nonlinear interaction, $g \sim 0$, the vortex solitons completely untangle, as evidenced in  Fig.~\ref{hawk2}  (by $t=21K$), when the internal energy in (\ref{energies}) satisfies $E_\text{int} \lll E_\text{kin}, E_\text{qu}$.
Surprisingly in 3+1 dimensions, the Poincar\'e recursion time for the GP equation (\ref{Gross_Pitaevskii_equation}) can be extremely short.
Furthermore, fast Poincare recurrence occurs only for simple initial conditions.

In a strictly unitary quantum algorithm, it is possible to observe recurrence because the numerical method is information preserving.  In a simulation in a finite-sized box with periodic boundary conditions, when kelvin waves excite phonons (incompressible energy changing to compressible energy) this is indeed acts like a  dissipative mechanism on the large scale and it gives rise to the characteristic $k^{-3}$ power law that we observe at large $k$.  However, the excited phonons cannot escape off to infinity since the simulation is in a finite-sized box.  At the point half way through one Poincare recurrence cycle, all the available phonon states are essentially filled and the quantum fluid can no longer transfer incompressible energy into these compressible modes.  It is at this point that the net transfer of energy reverses, and an inverse cascade is observed.  Although, we refer to the Poincare recurrence time as fast, this is a relative term.  On a small $512^3$ grid, one would have to run through over $\sim 20,000$ updates to see recurrence and this is a rather long simulation time. Previously the  computational resources where not available for do this kind of long run time, particular for pseudo-spectral algorithms that slowdown on large grids.  Furthermore, using complicated initial condition, such as the Taylor-Green profile \cite{nore:2644}, does not admit the shortest Poincare cycle time in the first place.  Finally, as far as we can tell,  tests of turbulent superflow in the case with $g$ very small have not been conducted.  This is why recurrence has not been previously observed.

\begin{figure}[bthp!]
\begin{center}
\includegraphics[width=3.4in]{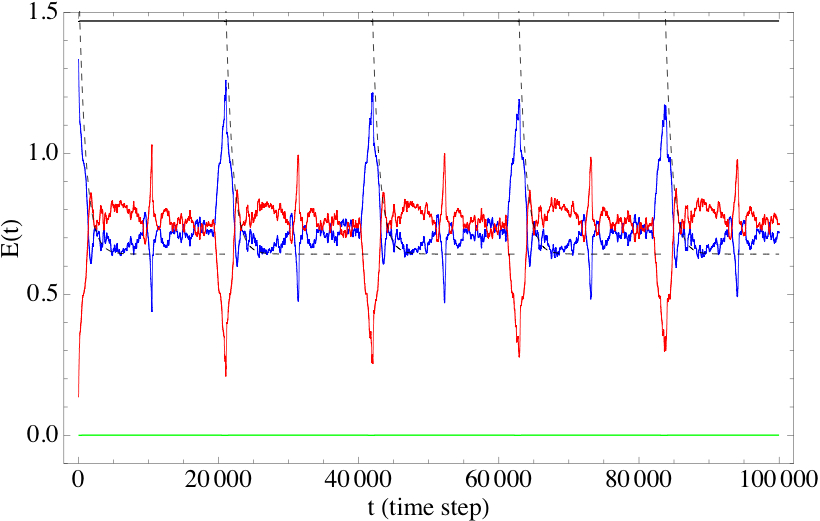}
\includegraphics[width=3.4in]{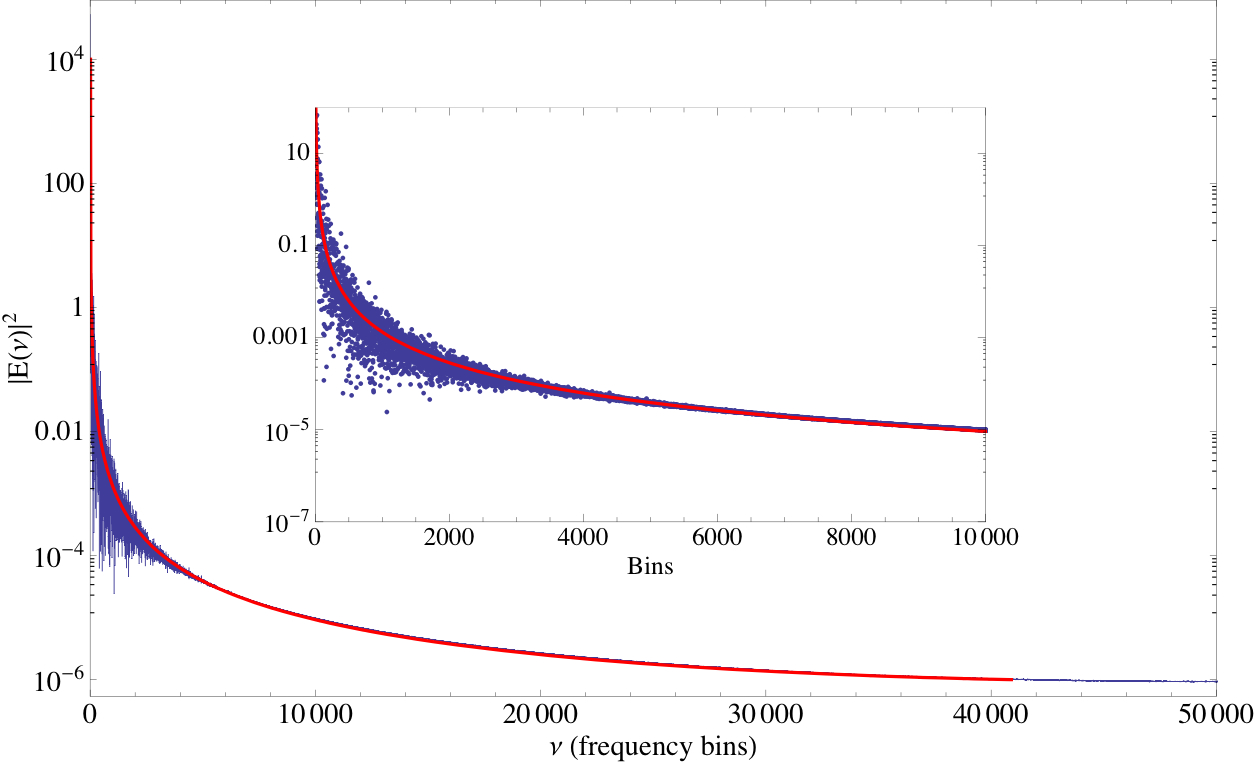}
\caption{\label{GPEnergies} \footnotesize  Top: The time evolution of the kinetic energy (blue), quantum energy (red), internal energy (green), and total energy (black line).  The total energy is conserved.  A recurrence of $t_\text{\tiny P}=20.9K$ time steps is determined from the envelop of the forward energy cascade (dashed). Bottom: Power spectrum indicates all frequencies are represented as $|E(\nu)|^2\sim [\nu(\nu_\text{\tiny max}-\nu)]^{-\frac{9}{4}}$ (red).
}
\end{center}
\end{figure}
To quantify this phenomenon, one can plot a time history of the kinetic, quantum and internal energy parts.
	In the $N=12$ simulation with $H_\text{int}(|\phi|^2)\sim 0$, the reverse cascade (absorption of sound waves) clearly repeats, cycling at $t_\text{\tiny P}\sim 21K$, the Poincar\'e recursion time for grid size $L=512$.    This recurrence time is clearly evidenced in the time evolution of the (rescaled) kinetic $E_\text{kin}$ and quantum $E_\text{qu}$ energies plotted in Fig.~\ref{GPEnergies}.
	
	The Poincar\'e recurrence theorem states that for Hamiltonian systems the solution trajectory passes arbitrarily close to the initial state provided the evolution is followed for a sufficiently long time.   While for certain maps in two spatial dimensions, like the Arnold Cat Map, the Poincar\'e recurrence time can be short, for nearly all Hamiltonian systems the recurrence time is so long as to be effectively infinite.  
There have been some analytical hints that the NLS equation in 1+1 dimensions could have a fast Poincare\'e recurrence time 
\cite{PhysRevLett.53.218}
--but this result was not expected to hold in three spatial dimensions.   From a series of quantum  simulations we see the Poincar\'e recursion time scales as  $L^2$ , where $L$ is the (linear) grid size. 
This arises from the inherent diffusive ordering of fluctuations, $\frac{\hbar}{m}\delta t \approx \delta x^2$.

\begin{figure*}[bthp]
\begin{center}
\includegraphics[width=1.7in]{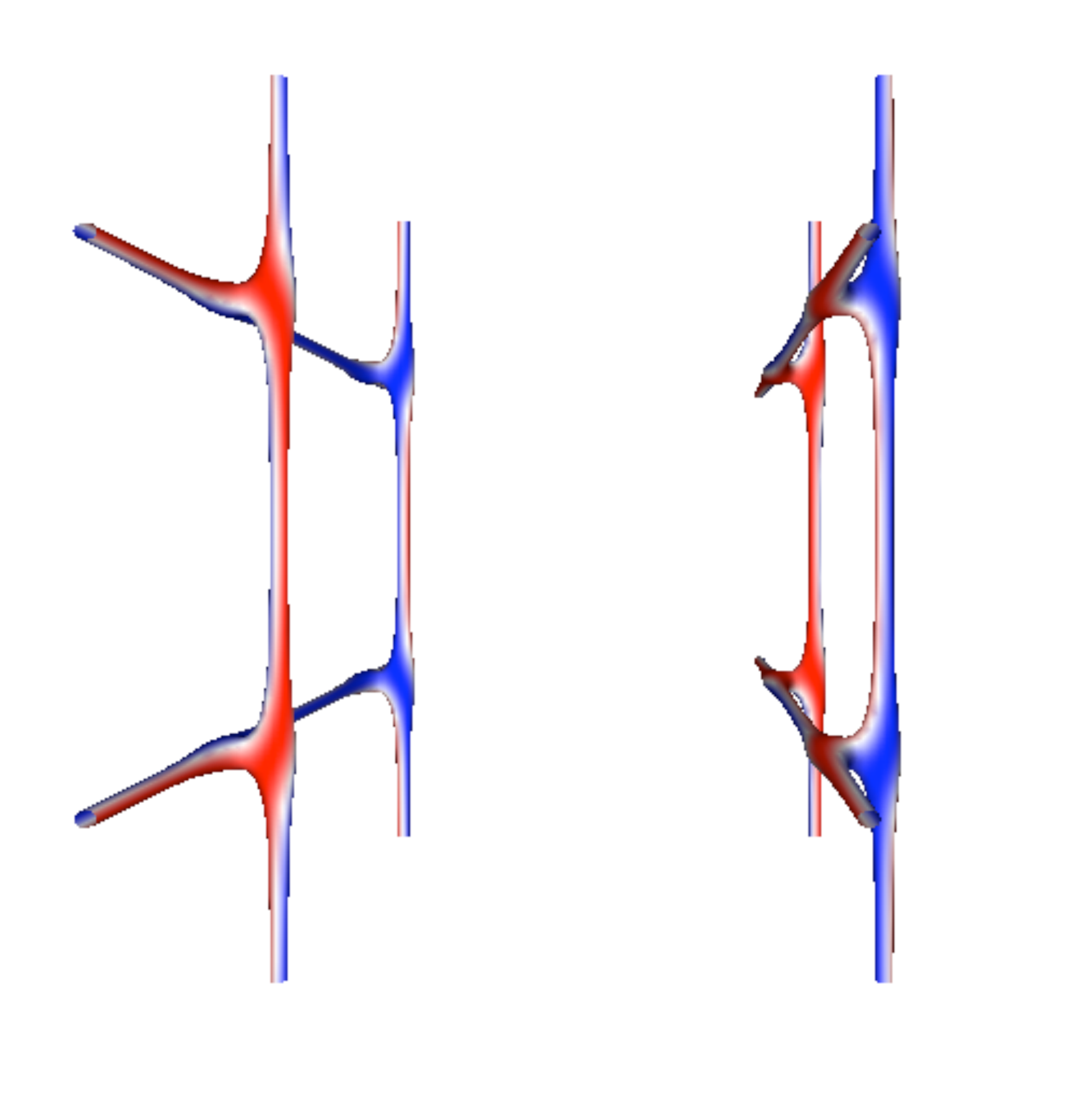}
\includegraphics[width=1.7in]{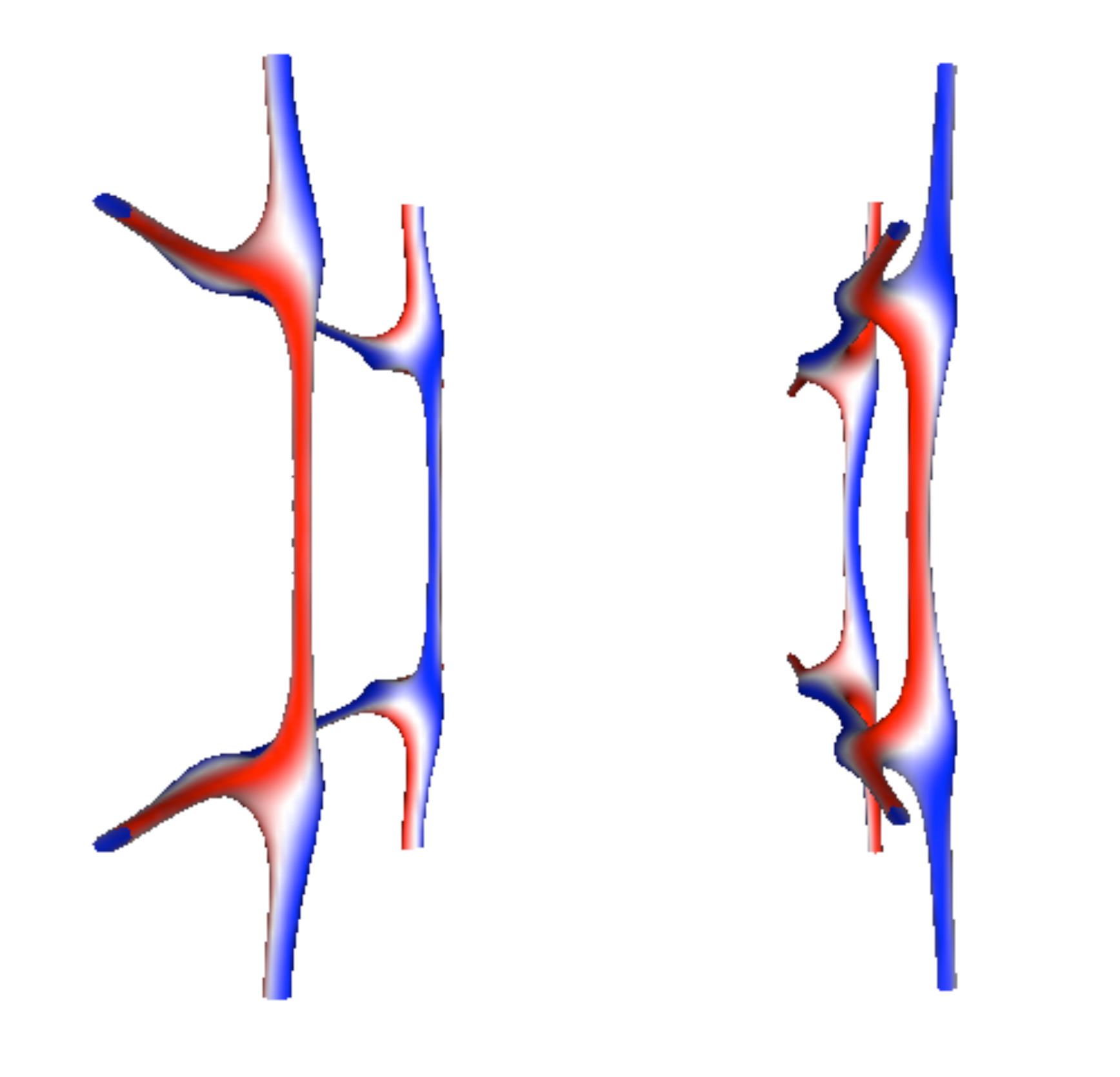}
\includegraphics[width=1.7in]{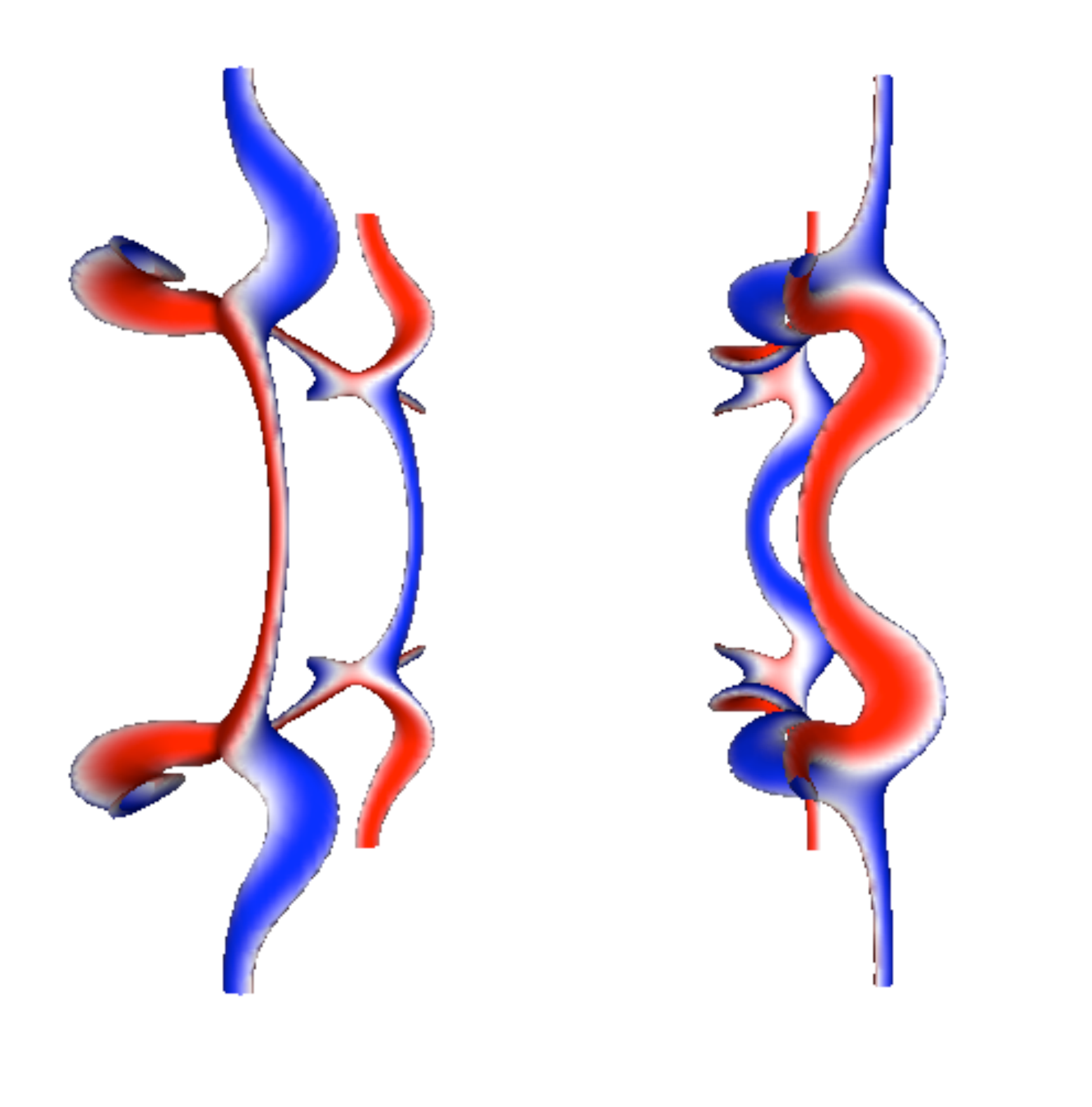}
\includegraphics[width=1.7in]{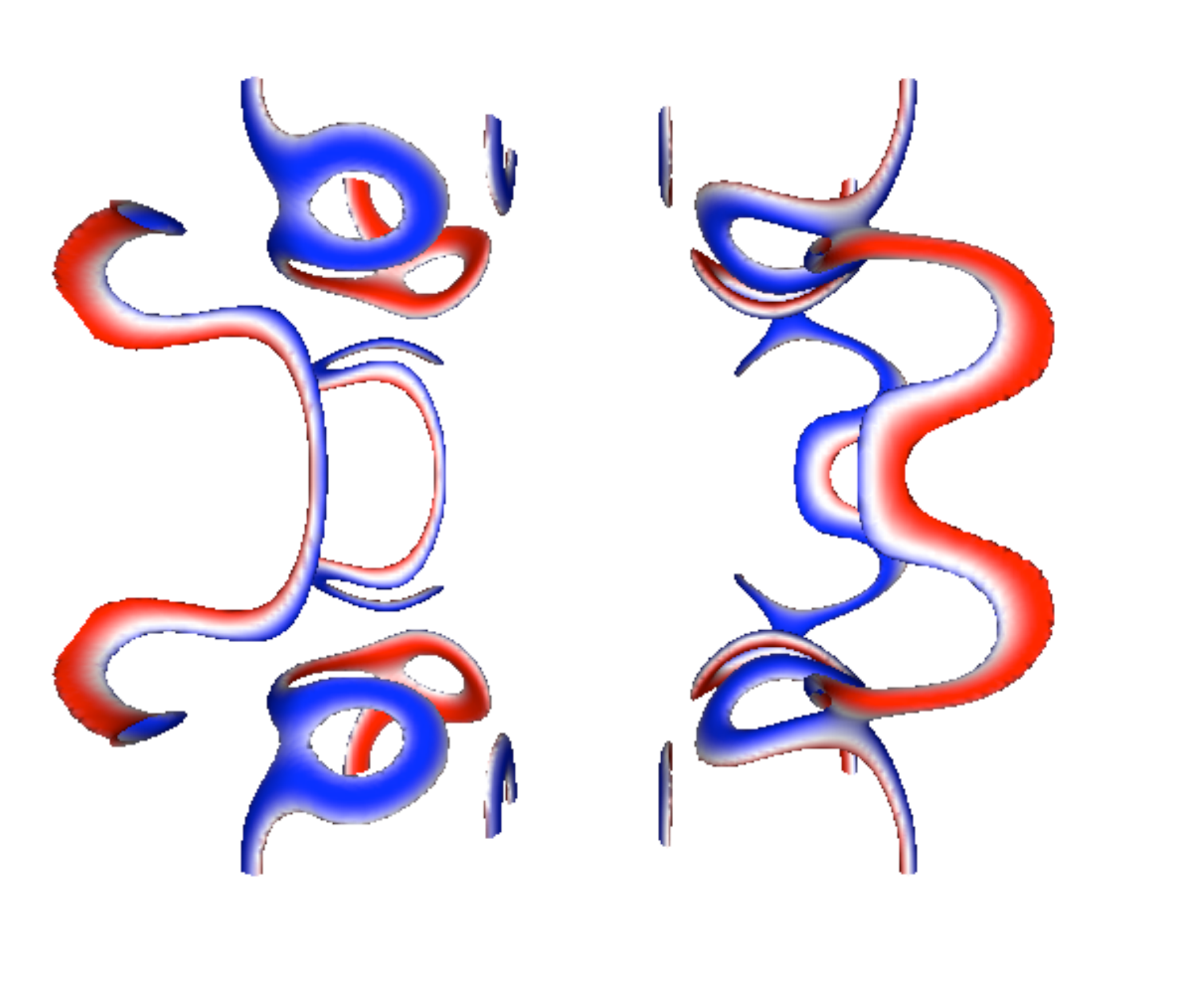}
\end{center}
\caption{\footnotesize  \label{twisting}
Kelvin waves seen as twisting of $(N=8)$ vortex filaments when $H_\text{int}(|\phi|^2)=|\phi|^2(1-  |\phi|^2)$.  In the $g\sim 0$ limit (non-interacting particles), there is a fast Poincar\'e recurrence time of $t_\text{\tiny P}=2020$.  For $g\approx 5$, vortices at the first few Poincar\'e cycles ($t = 0, t_\text{\tiny P}, 2t_\text{\tiny P}, 3t_\text{\tiny P}$) are plotted (top to bottom).  The Kelvin wave twisting in the vortices eventually completely breaks the fast Poincar\'e recurrence.  The highly tangled vortices, similar to that of Fig. 2, occurring between the Poincar\'e periods are not shown.  The asymmetry in the time is due to the broken symmetry of the initial condition.
}
\end{figure*}
%

\section{Isolating Kelvin waves}


We examined hydrodynamic-scale breaking of the quantum-scale time-reversal symmetry of the free system.
  $H_\text{int}$ breaks the Poincar\'e recursion and has a prominent effect on the dynamics of the quantized line vortices. 	Yet, it remains useful to chart the pathway to turbulent configurations at intervals demarcated by the Poincar\'e recurrence time of the free system. In the interacting GP limit,
 the fast recursion is broken by nonlinear twisting (Kelvin waves) riding on the originally linear vortices, $H_\text{int}$ successively twisting the filamentary centers every recursion period; see Fig.~\ref{twisting}.
The linear vortex tubes become tangled but at the free recurrence period they do not return to their original linear configuration. Instead, they become twisted and this twisting increases with more and more free Poincar\'e cycles.  

At very large times, the BEC manifests pure quantum turbulence, characteristic of nonlinear fluid behavior.
In this numerical simulation, $L=160$ and the smallness in the nonlinear interaction in (\ref{quantum_equation})  is set to $\frac{g\tau'}{\hbar} = 0.1$.   To sufficiently resolve the vortex core, the scale factor in the Pad\'e approximant
 is set to $a=0.05$. A convention of unity normalization is used ($\int  |\phi(x)|^2 dx^3 = 1$).
Poincar\'e recurrence in the $g\approx 0$ limit occurs at $t_\text{\tiny P} \simeq 2020 K$, and this time period is used to sample the wave function configurations of the  GP quantum system with $g\approx 5$.

\section{Conclusion}

A quantum lattice-gas algorithm for modeling hydrodynamic-scale BEC superfluid flow was presented.  The method accurately captures the dynamical behavior the superfluid even in the difficult case of fully developed quantum turbulence. Interacting vortex soliton lead to fluid instabilities in the unitary quantum fluid that causes an energy cascade in the regime of small wave numbers, leading to power law behavior $k^{-\frac{5}{3}}$ Kolmogorov turbulence in classical Navier-Stokes fluids.  However, at larger wave numbers (approaching the scale of the vortex cores), the spectrum of incompressible kinetic energy transitions to an alternate power law unique to quantum fluid flow, than finally to a $k^{-3}$ for very large wave numbers (corresponding to fluid dynamics internal to the vortex cores).  This power-law behavior has recently been observed in experimental observations of superfluid Helium II \cite{paoletti:154501}.

For weakly nonlinear BEC superfluids, it is possible to observe fast Poincar\'e recurrence where at each recurrence time, Kelvin waves are observed to emerge with greater amplitude at each recurrence time, {\it viz.}, vortex solitons that are successively twisted.

The quantum algorithm tested here hopefully will be implemented on a large quantum computer someday, if that device becomes available.  At that time, the many-body quantum simulation problem can be directly addressed numerically using the quantum complexity inherent to the quantum computer.  This would include interference effects across the $n$-body sectors of the Hilbert space caused by the nonlinear particle-particle interactions, whereas in our present supercomputer-based simulation we have restricted ourselves to modeling a one-component BEC superfluid in the one-body sector.

\section{Acknowledgement}

We thank S. Ziegeler, M. Soe, and J. Carter for help with graphics and a parallel version of our code run at the Air Force Research Laboratory, with scalings up to $5120^3$.  
Finally, we would also like to thank the Naval Oceanographic Office and the Air Force Research Laboratory's High Performance Computing Major Shared Resource Centers for  supercomputer allocations.  JY thanks Seth Aubin and John Delos for discussions about BEC experiments.  JY would also like to thanks Norman Margolus for helpful discussions.


\bibliographystyle{apsrev}

\end{document}